\pdfoutput=1
\documentclass[aps,prl,10pt,twocolumn,showpacs,superscriptaddress]{revtex4-1}
\usepackage{amsmath}
\usepackage{latexsym}
\usepackage{pgfplots}
\usepackage{rotating}
\usepackage{amssymb}
\usepackage{bm}
\usepackage{graphics,epstopdf}
\usepackage{color}\usepackage{amsmath}

\usepackage[usenames,dvipsnames,svgnames]{pstricks}
\usepackage{epsfig}
\usepackage{pst-grad} 
\usepackage{pst-plot} 

\usepackage{newlfont}
\usepackage{amssymb}
\usepackage{amsfonts}
\usepackage{amsmath}
\usepackage{mathrsfs}
\usepackage{amsthm}
\usepackage{graphicx}
\usepackage{epsfig}
\usepackage{color}
\usepackage{bm}
\usepackage[breaklinks]{hyperref}
\usepackage{tikz}
\newcommand{\ket}[1]{|{#1}\rangle}
\newcommand{\bra}[1]{\langle{#1}|}

\usepackage{times}

\begin{document}
	\title{Coherence of Purification}

	\author{Arun Kumar Pati}
	\email{akpati@hri.res.in}
	\affiliation{Quantum Information and Computation Group, Harish-Chandra Research Institute, HBNI, Allahabad, 211019, India}
\author{Long-Mei  Yang}
	\affiliation{School of Mathematical Sciences, Capital Normal University, Beijing 100048, China}
	
\author{Chiranjib Mukhopadhyay}
\affiliation{Quantum Information and Computation Group, Harish-Chandra Research Institute, HBNI, Allahabad, 211019, India}	
	\author{Shao-Ming Fei}
	\affiliation{School of Mathematical Sciences, Capital Normal University, Beijing 100048, China}

	\author{Zhi-Xi Wang}
	\affiliation{School of Mathematical Sciences, Capital Normal University, Beijing 100048, China}
	
	\date{\today}

	\begin{abstract}
We introduce a quantity called the coherence of purification which can be a measure of total quantumness for a single system. We prove that coherence of purification is always more than the coherence of the system. For a pure state, the coherence of purification is same as the relative  entropy of coherence. Furthermore, we show that the difference in the coherence of purification of a quantum state before and after the dephasing can capture
residual quantumness.
In addition, we show that the entanglement of purification that can be created by incoherent operation between two subsystems is upper bounded by the coherence of purification of the original system. We show that the coherence of purification is a resource in quantum state discrimination process. In the absence of coherence, entanglement and quantum correlations, the coherence of purification may play an important role in quantum information processing tasks.
	\end{abstract}

	\maketitle
		
	
	\section{Introduction}
The concept of superposition of quantum states for quantum systems is one of the most important feature that makes  quantum system dramatically different than classical system. Superposition gives rise to coherence and  recently, there has been a considerable effort devoted towards developing resource theory of quantum coherence \citep{plenio, aberg, winter, singh, adesso_colloquium,theurer, sreetama, Uttam, swapan, Mani_CP, Bu_CP, kai}. This is also a primitive resource which is central to other non-classical resources and
 plays a key role in quantum foundation \cite{bera}, quantum thermodynamics \citep{lostaglio1, lostaglio2, varun}, and various other directions \citep{tan, magic, mile, ma, united, piani, symmetric}.  The resource theory of coherence, developed in Ref. \cite{plenio}, provides a set of conditions on a real valued function of quantum states for it to be a bona fide quantifier of quantum coherence. This resource theory is based on the set of incoherent operations as the free operations and the set of incoherent states as the set of free states. The set of incoherent states and the set of incoherent operations depend crucially on the reference basis.

 When we think of composite quantum systems, they can contain variety of quantumness that are usually not seen in the classical world. Last several years have witnessed some of these non-classical features for bipartite and multipartite quantum systems such as entanglement \citep{ent_rmp}, discord \citep{hv, oz, modi}, work deficit \citep{oppenheim, ujjwal, ujjwal2}, measurement induced non-locality \citep{min}, non-local advantage of  quantum coherence \citep{deba}, and many more which are useful resources in quantum information processing. Given a single quantum system we know that it can have some coherence which arises from the linear superposition principle. However, it is still not fully clear what are various quantumness measures beyond coherence that one can associate with a single quantum system, although various other resource theories, e.g., purity \citep{purity}, or imaginarity \citep{hickey_gour} have been put forward.
	
We introduce a new quantity, formally analogous to the entanglement of purification \citep{eop}, called as the coherence of purification for a density operator. This is not a measure of quantum coherence, in general, as it does not satisfy properties of a coherence monotone. It is the coherence of the purified state of the density operator with minimum taken over all possible purifications.
Thus, it is an inherited quantumness from the Church of the large Hilbert space. The coherence of purification is always larger than the quantum coherence and hence this can capture total quantumness for a single quantum system. In this sense, the coherence of purification may capture quantumness beyond coherence. The coherence of purification
for a pure state is same the coherence of the state itself, as the state is already a pure state. Therefore, for a
pure state there is nothing beyond quantum coherence.  If we take a
dephased version of the density operator it may still have non-zero coherence of purification. This means that
an incoherent state in a given basis can have finite amount of coherence of purification. This, then motivates us to define a residual quantumness as the difference between the coherence of purification of the original state and the dephased state. This is because, even after the system looses all its coherence there can be
some quantumness left inside.
Furthermore, we show how to convert the coherence of purification to entanglement of purification using only incoherent operation, thus, extending the notion of inter convertibility between different resources. Also, we show that for bipartite states the coherence of purification can be more than the entanglement of purification (a measure of total correlations) as well as entanglement of formation. As a concrete application, we illustrate how coherence of purification can be regarded as a resource for quantum state discrimination as well as an indicator of the success of the state discrimination. We hope that the coherence of purification may play an pivotal role in single as well as composite systems where
quantum coherence, entanglement, quantum discord and other quantum correlations are absent.

	\section{Resource Theory of Coherence }
Resource theory of coherence needs free states and free operations which are essentially
 the set of incoherent states and incoherent
	operations, respectively. Since coherence is basis dependent, we can fix a basis. Let us consider the computational basis,
	$ \ket{i} ( i = 0,1, \ldots N) $ in Hilbert space ${\cal H}$, with $ N =$  dim ${\cal H} $.
	The diagonal density matrices in this basis are incoherent states and expressed as
	\begin{equation}
	\label{eqa}
	\delta=\sum_{i} \delta_{ii}\ket{i}\bra{i}.
	\end{equation}
	The set of incoherent states is represented by $\mathcal{I}$. The operations which keeps all incoherent states incoherent, are called incoherent operations.
	
	The quantification of resource is an important aspect for its physical implications. Before going to define the measure of coherence of purification, we state here  the basic requirements for a valid measure of coherence. A valid measure of quantum coherence satisfies \cite{plenio}: (i) Coherence vanishes for all incoherent state, $C(\delta)=0$ for all  $\delta\in\mathcal{I}$, (ii)
Coherence should not increase under mixing of states, i.e., $\sum_ip_iC(\rho_i)\geq C(\sum_ip_i\rho_i)$, (iii)
Monotonicity under incoherent completely positive and trace preserving (CPTP) maps, $\Phi$: $C(\rho)\geq C(\Phi[\rho])$, and (iv)  Monotonicity under selective incoherent operations on average $C(\rho)\geq \sum_{i} p_n C(\rho_n)$, where $p_n=  Tr[{K_n}\rho K_n^{\dagger}]$, and $\rho_n=\frac{1}{p_n}{K_n}\rho K_n^{\dagger}$ with $\{ K_n \} $ is Kraus decomposition of $\Phi$.
	
In the present paper, we will make use of the relative entropy of coherence for a state $\rho$ which is defined as
\begin{align}
C_R(\rho) = min_{\sigma \in {\cal I} }  S(\rho ||\sigma),
\end{align}
where ${\cal I}$ denotes the set of incoherent states. The relative entropy of coherence admits the closed expression and is given by
$C_R(\rho)  = S(\Delta[\rho] - S(\rho)$, where $\Delta[\rho]$ is the dephased density operator in the basis $\{\ket{i} \}$. This is also same as the distillable coherence of the state $\rho$ which is the optimal rate for asymptotic
extraction of maximally coherent qubits via incoherent operations. Throughout this paper, we use the notion of relative entropy of coherence to define
the coherence of purification.

\section{Coherence of Purification}

A quantum system $A$ is associated with a Hilbert space ${\cal H}_A $ with dimension $ |A| := dim({\cal H}_A)$. Let ${\cal L}(A)$
represent the set of all linear operators on ${\cal H}_A $. We denote the set of quantum states on the Hilbert space ${\cal H}_A $ by ${\cal B}(A) $.
The physical system in a state $\rho$  with subscript $A$ indicates $\rho_A \in {\cal B}(A)$. If the system is a pure state,
we use $\Psi$ to represent the pure state density matrix $\ket{\Psi} \bra{\Psi}$.
Given a $\rho_A \in {\cal B}(A)$, a purification of $\rho_A$ is a pure state $\Psi_{AB} \in  {\cal B}(AB)$ with $\rho_A= Tr_B \Psi_{AB}$.
Purification of a given density operator is not unique and can have infinite number of purifications.

First, we define the coherence of purification for a density operator $\rho$.

{\bf Definition:}
Let $\rho = \rho_A$ be a density operator on ${\cal H}_A$. Let $\ket{\Psi}_{AB} $ is the purified state of $\rho$ with $ \rho_A = Tr_B [\ket{\Psi}_{AB} \bra{\Psi}] $.
The coherence of purification for a single system in a state $\rho$ is defined as
 \begin{equation}
	\ C_{P}(\rho)= min_{B} C_R (\Psi_{AB}),
	\end{equation}
where  we have denoted $\Psi_{AB} = \ket{\Psi}_{AB} \bra{\Psi}$ and  the relative entropy of coherence is defined as
	\begin{equation}
	\label{eqd}
	\ C_{R} (\Psi_{AB}))=S(\Delta[\Psi_{AB}] ).
	\end{equation}
If $\rho=  \sum_i p_i \ket{\psi_i}  \bra{\psi_i}$, with $\ket{\psi_i}'s$ being non-orthogonal, then the purified state can be written as
  $\ket{\Psi}_{AB} = \sum_i \sqrt{p_i} \ket{\psi_i} \ket{i} $. Since there are infinite number of purifications possible for a given density operator, the coherence of purification can be expressed as
\begin{equation}
	\ C_{P}(\rho)= min_{U_B} C_R (I_A \otimes U_B \ket{\Psi}_{AB} \bra{\Psi}  I_A \otimes U_B^{\dagger}).
	\end{equation}

In the sequel, we give explicit expressions for these quantities. Consider
 a state $\rho=  \sum_i p_i \ket{\psi_i}  \bra{\psi_i}$ with  the purified state as  $\ket{\tilde{\Psi}} = (I \otimes U_B)\ket{\Psi}_{AB} = \sum_i \sqrt{p_i} \ket{\psi_i} U_B\ket{i}=
  \sum_i \sqrt{p_i} \ket{\psi_i} \ket{b_i}  $, where $\ket{b_i} = U_B \ket{i}$.
Let us define the quantum coherence of the purified state using the basis  $\{ \ket{i}_A \ket{b_j}_B \}$. The dephased state in this basis is then given by
\begin{align}
\Delta[\ket{\tilde{\Psi}} \bra{\tilde{\Psi}}]
 = \sum_{ij} g_{ij} \ket{i}\bra{i} \otimes \ket{b_j}\bra{b_j}
\end{align}
where $g_{ij} = p_j f_i^{(j)}$, $f_i^{(j)} = |\bra{i}\ket{\psi_j}|^2 $ with $\sum_{ij} g_{ij} =1$ and $\sum_i |\bra{i}\ket{\psi_j}|^2 =1$ for all $j$. Therefore, the coherence of purification for any density operator in these basis is
\begin{align}
 C_P(\rho) &= min_{U_B} C_R ({\tilde {\Psi}} ) = - \sum_{\mu} g_{\mu} \log g_{\mu}, (\mu =ij) \nonumber\\
& =  H(p_j) + \sum_j p_j H(f_i^{(j)}).
\end{align}
Notice that we are using the basis $\{ \ket{i}_A \ket{b_i}_B \}$ to define the coherence, this helps to avoid the minimization. However, if we define coherence
in canonical basis such as $\{ \ket{i}_A \ket{i}_B \}$, then we need to minimize. Thanks to the Schmidt decomposition theorem, the coherence of
purification can be evaluated without minimization.

\section{Properties of Coherence of Purification}

Here, we prove some useful properties of the coherence of purification. We first prove that $C_P$ is always larger than $C_R$. Then, we prove its monotonicity under
dephasing and genuinely incoherent operations. Thereafter, we prove the monotonicity of $C_P$ under selective incoherent operation. 
Though, the Coherence of purification is not a monotone under general incoherent channel, proving that it is monotonous under a specific subset of incoherent operations, shows a handy link with the usual resource theory of quantum coherence. Finally, we show that it 
satisfies continuity, that is, if two states are sufficiently close, their coherences of purifications differ only by a small amount.  We further obtain bounds of coherence of purification in terms of the mixedness and the coherence cost of the relevant quantum state. Finally, we put forward propositions which seek to show the connection between the entanglement of purification and our formulation of coherence of purification for any bipartite quantum states.

Now, we will show that the coherence of purification  satisfies following properties:\\

{\bf Proposition 1:}  For any state $\rho$, it holds that $\ C_{P}(\rho) \ge \ C_{R}(\rho)$. The equality holds if $\rho$ is a pure state.

{\bf Proof:}  For any bipartite state $\rho_{AB}$, the relative entropy of coherence satisfies
$C_R(\rho_{AB} ) \ge  C_R(\rho_{A} ) + C_R(\rho_{B} ) $. Let us assume that $\Psi_{AB}$ is optimal purification, then we have  $\ C_{P}(\rho)=  C_R (\Psi_{AB} )  \ge  C_R(\rho_{A} ) + C_R(\rho_{B} ) $. Hence, $\ C_{P} (\rho) \ge \ C_{R}(\rho)$. When $\rho$ itself is a pure state then the optimal purification can be  chosen such that
$C_R(\rho_B) = 0$ and hence  $\ C_{P}(\rho) = \ C_{R}(\rho)$.

This shows that the amount of quantum coherence
is always less than or equal to the total quantumness contained in a state. This is in accordance with our
intuition that for a mixed state there is some quantumness beyond coherence.

{\bf Proposition 2:}
Coherence of purification satisfies monotonicity under complete dephasing maps, i.e.,  $C_P(\rho) \geq C_P(\Delta[\rho])$, where
$\Delta[\rho] = \sum_k \Pi_k \rho \Pi_k$ with $\Pi_k = \ket{k}\bra{k}$.\\

{\bf Proof:}
The dephased state $\Delta[\rho]  = \sum_k \lambda_k \ket{k}\bra{k}$, where
$\lambda_k = \sum_i p_i |\bra{k} \psi_i \rangle|^2$. Consider the purification of $\rho^D$ as
$\ket{\tilde{\Phi}} = (I \otimes U_B)\ket{\Phi}_{AB} = \sum_i \sqrt{\lambda_i} \ket{i} \ket{b_i}$. The coherence of purification for $\Delta[\rho]$ is given by
\begin{align}
 C_P(\Delta[\rho]) = min_{U_B} C_R ({\tilde {\Phi}} ) = min_{U_B} S ({\tilde {\Phi}}^D ) =  - \sum_k \lambda_k \log \lambda_k,
\end{align}
where the coherence is again defined in $\{ \ket{i}_A \ket{b_i}_B \}$ basis.

 Note that the relative entropy of coherence $C_R (\rho_{AB} )$ satisfies
\begin{align}
C_R (\rho_{AB} )  \ge C_R(\rho_A) + C(\rho_B) + max \{ D_{A \rightarrow B}, D_{B \rightarrow A} \},
\end{align}
where $D_{A \rightarrow B}$ and $D_{B \rightarrow A}$ are the left and right sided quantum discords, respectively. Since $\rho_{AB} = \tilde{\Psi}_{AB}$ and for pure state both discords are same, i.e.,we have $D_{A \rightarrow B} = D_{B \rightarrow A} = S(\rho_A) = S(\rho_B)$. Further, in the basis $\{ \ket{i} \}$ we have
$C_R(\rho_A)=  -  \sum_k \lambda_k \log \lambda_k  - S(\rho_A) $  and in the basis $\{ \ket{b_i} \}$ we have
$C_R(\rho_B)= 0$. Therefore, this results  in

\begin{align}
C_P(\rho) = C_R ({\tilde{ \Psi_{AB} }} )  \ge C_P(\Delta[\rho])
\label{diag_is_less}
\end{align}

Now, let us briefly recapitulate that entanglement of purification is monotonic under local operations. However, the coherence of purification is non-monotonic under arbitrary incoherent operations, an example being the completely depolarizing channel. Therefore, it is natural to wonder whether there exists any subset of incoherent operations, just as local operations are a subset of separable operations, for which the coherence of purification is monotonic. Below, we answer this question in the affirmative for a particular class of incoherent operations known as \emph{genuinely incoherent operations} (GIO) \citep{gio}. These are defined as incoherent operations with the added constraint that $\Lambda_{GIO}$ preserves every diagonal state $\sigma_{\text{diag}}$, that is
\begin{equation}
\Lambda_{GIO} (\sigma_{\text{diag}}) = \sigma_{\text{diag}} \forall \sigma_{\text{diag}}
\end{equation}

{\bf Proposition 3:}  Coherence of purification $C_{P}$ is monotonic under Genuinely Incoherent Operations (GIO), that is for every GIO channel $\Lambda_{GIO}$ acting on a quantum state $\rho_A$,

\begin{equation}
C_{P} (\rho_A) \geq C_{P} (\Lambda_{GIO} (\rho_A))
\label{upper}
\end{equation}

\textbf{Proof: } It is known \citep{gio} that the Kraus operators $\lbrace K_n \rbrace$ corresponding to the genuinely incoherent operations are diagonal in the relevant basis $\lbrace | i \rangle \rbrace$, i.e.,

\begin{equation}
K_{n} = \sum_{i} \lambda_{i}^{(n)} |i\rangle \langle i|
\end{equation}

\noindent Now, let us first show that the dephased version of a state is identical to the dephased version of the state after undergoing a GIO, that is, if $\Delta$ is the dephasing map, then \begin{equation}
\Delta[\Lambda_{GIO} (\rho)] = \Delta[\rho]
\end{equation}

\noindent To show this, let us note that the dephased version of the state which has already undergone the GIO channel above is given by $\Delta[\Lambda_{GIO} (\rho)] = \sum_{j,n} \langle j | K_n \rho K_n^{\dagger}|j\rangle |j\rangle\langle j| = \sum_{j,n} |\lambda_{j}^{(n)}|^2 \langle j|\rho|j\rangle |j\rangle \langle j| =  \sum_{j} \langle j|\rho|j\rangle |j\rangle \langle j| = \Delta [\rho]$. In the penultimate step, we used the normalization property of the Kraus operators, i.e., $\sum_{n} K_{n}^{\dagger} K_n = \mathbb{I}$.

\noindent Now, let $|\Phi\rangle_{AB}$ be the purified state corresponding to $\Lambda_{GIO} (\rho_A)$. Then, the coherence of purification of $\Lambda_{GIO} (\rho_A)$ is given by $C_{R} ( \Phi_{AB}) = C_{R} (\Lambda_{GIO} (\rho_A)) + C_{R} (\rho_B) + D (\Phi_{AB})$, where $D$ is the symmetric discord, which is replaced by the entanglement entropy for pure states. Because the entropy of any subsystems of a pure state are equal, the second and the third term both equals the von Neumann entropy of $\Lambda_{GIO} (\rho_A)$, while $C_{R} (\Lambda_{GIO} (\rho_A)) = S(\Delta(\Lambda_{GIO} (\rho_A)))$.

\noindent However, by the earlier result we proved, $C_{R} (\Lambda_{GIO} (\rho_A)) = S(\Delta(\Lambda_{GIO} (\rho_A))) = S(\Delta(\rho_A)) =C_{P} (\rho_{A}^D) $.  Now, by Eq.~\eqref{diag_is_less}, $C_{P} (\Delta (\rho)) \leq C_{P} (\rho)$. Thus,  $\Lambda_{GIO} (\rho_A) \leq C_{P} (\rho_A)$. This concludes the proof. \qed

This is analogous to the case of entanglement of purification which is monotonic under local operations.

{\bf Proposition 4:}
Coherence of purification cannot increase on average under selective incoherent operations, i.e., it satisfies 

 $$C_P(\rho)\geq \sum_{n} p_n C_P(\rho_n), $$
 where $p_n=  Tr[{K_n}\rho K_n^{\dagger}]$,
	    and $\rho_n=\frac{1}{p_n}{K_n}\rho K_n^{\dagger}$ with $\{ K_n \} $ is Krauss decomposition of $\Phi$.

{\bf Proof:} Let $\ket{\Psi}_{AB} $ is the optimal pure state for $\rho$.  When $\rho \rightarrow
\rho_n=\frac{1}{p_n}{K_n}\rho K_n^{\dagger}$, it is equivalent to $\ket{\Psi}_{AB} \rightarrow \ket{\Psi_n}_{AB} =
\frac{1}{p_n} (K_n \otimes I ) \ket{\Psi}_{AB} $.  Now, using the monotonicity of $C_R(\Psi_{AB})$ under selective incoherent operation, we have 
$C_P(\rho) = C_R(\Psi_{AB}) \ge \sum_n p_n
C_R(\ket{\Psi_n}_{AB}  \bra{\Psi_n} ) \ge \sum_n p_n C_P(\rho_n) $. The last inequality follows from the fact
that $\ket{\Psi_n}_{AB}$ is one possible purification for $\rho_n$, but may not be the optimal one.

 It should be mentioned that the coherence of purification is neither convex nor concave, i.e., it does not satisfy the convexity property.
 For example, we could have a incoherent mixed state $\rho = p_0 \ket{0}\bra{0} + p_1 \ket{1}\bra{1}$ and for this we do not have
 $C_P(\rho) \le p_0 C_P( \ket{0}\bra{0}) + p_1 C_P(\ket{1}\bra{1})$. Because, this suggests that $C_P(\rho) = -p_0 \log p_0 - p_1 \log p_1$ and $RHS = 0$,
 thus violating convexity. Similarly, one can also find example where it does satisfy convexity.  Also, the coherence of purification can increase or decrease under
 incoherent channel. For example, consider a randomizing channel that transforms any state $\rho \rightarrow {\cal E}(\rho) = \frac{I}{d}$. In this case, take $\rho$
 as an incoherent state  $ \sum_i p_i \ket{i}\bra{i}$. Then, $C_P(\rho) = - \sum_i p_i \log p_i$. However, the coherence of purification for $\frac{I}{d}$ is
 $\log d$, i.e., under randomizing channel it can increase.

Since the coherence of purification does not satisfy the monotonicity and convexity akin to coherence measure, this is not a quantifier of quantum coherence. In fact, as this aimed to be, it captures something extra compared to quantum coherence.

{\bf Proposition 5 :} Coherence of purification $C_{P}$ is continuous, that is, if two quantum states $\rho_A, \sigma_A$ belong to the same Hilbert space $\mathcal{H}_A$, and the Uhlmann fidelity between them is $\mathcal{F}$, then the following inequality holds.

 \begin{equation}
 C_{P} (\sigma) - C_{P} (\rho) \leq 2 \sqrt{1 - \mathcal{F} (\rho, \sigma)} \log d + h(\sqrt{1 - \mathcal{F} (\rho, \sigma)}) ,
 \end{equation}

\noindent where $h$ is the Binary entropy, and $d$ is the dimension of the Hilbert space.

{\bf Proof :} We shall assume without loss of generality that $C_{P}(\rho) \leq C_{P} (\sigma)$. Now, let us assume the purifications of $\rho$, and $\sigma$, respectively denoted as $|\Psi\rangle$, and $|\Phi\rangle$, are the ones that optimise the Uhlmann Fidelity $\mathcal{F}$ between these quantum states. These are of dimension $d^2$. Now, let us further assume that the optimal purification $|\Psi ' \rangle$ corresponding to coherence of purification of $\rho$ is given by a unitary transformation $U$ on $|\psi\rangle$. The same untary applied on the purification of $\sigma$ will produce a state $|\Phi' \rangle$, which will, in general not be equal to the optimal pure state $|\xi\rangle$ for coherence of purification of $\sigma$. However, since $|\Phi'\rangle$, and $|\Psi'\rangle$ are unitarily connected, they also share the same optimal Uhlmann fidelity.

\noindent Thus, $C_{P} (\sigma) - C_{P} (\rho) \leq C_{R} (|\Phi'\rangle \langle  \Phi'|) - C_{R} (|\Psi'\rangle \langle  \Psi'|) = S[\Delta (|\Phi'\rangle \langle  \Phi'|)] - S[\Delta (|\Psi'\rangle \langle  \Psi'|)]$.

\noindent Now, by Fannes' inequality \citep{fannes}, $S[\Delta (|\Phi'\rangle \langle  \Phi'|)] - S[\Delta (|\Psi'\rangle \langle  \Psi'|)] \leq 2 T(\Delta (|\Phi'\rangle \langle  \Phi'|), \Delta (|\Psi'\rangle \langle  \Psi'|)) \log d^2  + h(T)$, where $h$ is the Binary entropy. Now, by Ruskai's proof of monotonicity of trace distance under CPTP maps \citep{ruskai}, $T(\Delta (|\Phi'\rangle \langle  \Phi'|), \Delta (|\Psi'\rangle \langle  \Psi'|)) \leq T(|\Phi'\rangle, |\Psi'\rangle)  = \sqrt{1 - \mathcal{F} (\rho, \sigma)}$.  Using this expression, we finally get the desired inequality

\begin{equation}
C_{P} (\sigma) - C_{P} (\rho) \leq \sqrt{1 - \mathcal{F} (\rho, \sigma)} \log d^2 + h(\sqrt{1 - \mathcal{F} (\rho, \sigma)}) \nonumber
\end{equation} \qed

Now, let us try to bound the coherence of purification from above. To this end, we have to  consider a coherence measure. A well known resource measure is the so called \emph{resource cost}, which quantifies the  rate at which pure coherent states have to be consumed to form a mixed state $\rho$. For coherence, the coherence cost equals the coherence of formation $C_f$ \citep{winter}. We show below, that the coherence of purification is bounded above by the sum of the coherence of formation and the mixedness in terms of von-Neumann entropy of the quantum state.

{\bf Proposition 6:}  For any state $\rho_A$, the coherence is purification $C_{P}$ is non-trivially bounded above by the sum of the coherence of formation $C_{f}$ and the von Neumann entopy $S$ of the state, that is,

\begin{equation}
C_{P} (\rho_A) \leq C_{f} (\rho_A) + S (\rho_A)
\label{upper}
\end{equation}

{\bf Proof: } Assume the optimal pure state decomposition in the context of the process of formation corresponding to the mixed state  $\rho_A$  is written as $\rho_A = \sum_{i} p_{i} |\psi_i \rangle_A \langle \psi_i \rangle $. Thus, the coherence of formation equals $C_F = \sum_{i} p_{i} C_R (|\psi_i \rangle_A) \geq C_R \left( \sum_{i} p_{i} |\psi_i \rangle_A \langle \psi_i | \right)$, where the last inequality follows from the convexity of relative entropy of coherence. Now, a generally non-optimal purification of the above state $\rho_A$ is the pure state $|\Psi \rangle_{AB} = \sum_i \sqrt{p_i} |\psi_i \rangle_A \otimes |i \rangle_B$. The coherence of purification $C_P (\rho_A)$ is thus upper bounded by $C_R \left( \sum_i \sqrt{p_i} |\psi_i \rangle_A \otimes |i \rangle_B \right)$. The probability vector corresponding to the dephased version of this state is given by $f_k = \text{diag} \lbrace p_n |\langle \psi_n | m \rangle|^2 \rbrace_{m,n}$, whose Shannon entropy  $H(f_k), k =(m,n)$ is identical to the the Shannon entropy of the dephased mixed state formed by the dephased version of the original state $\rho_A$.  Combining the above facts, we obtain the requisite relation $C_{P} (\rho_A) \leq  S(\rho_A)  + C_f (\rho_A) \qed$

\section{Residual quantumness}
Consider a quantum state in a mixed state $\rho$ and let us perform a complete projective measurement on
the system. After measurement the state changes as $\rho \rightarrow \sum_k \Pi_k \rho \Pi_k =  \Delta[\rho]$. The resultant density operator will be an incoherent one in the measured basis. However, the coherence of purification of the original and the dephased states are in general different. After the dephasing operation, the state looses all its coherence, yet its coherence of purification can be non-zero. We propose
that the coherence of the purification of the state before and after the complete measurement may capture
residual quantumness $Q_R$, i.e.,

\begin{align}
Q_R(\rho) = C_P(\rho) - C_P(\Delta[\rho]).
\end{align}
Since the coherence of purification cannot increase under dephasing operation, we have $C_P(\rho)  \ge  C_P(\Delta[\rho])$ and hence $Q_R(\rho)$ is positive.
Therefore, the residual quantumness in this basis is given by

\begin{align}
Q_R(\rho) =    \sum_k \lambda_k \log \lambda_k - \sum_\mu g_\mu \log g_\mu.
\end{align}

This shows that even if we choose an arbitrary basis for the ancilla system, any mixed state can contain some residual quantumness. However, for a pure state the
residual quantumness is identically zero. Therefore, for a pure state there is no quantumness beyond coherence and no residual quantumness.

We now work out an explicit example of calculating the coherence of purification for the family of qubit states $\rho_A = p |+\rangle\langle +| + (1-p) |-\rangle \langle - |$. Then a possible purification is $|\psi\rangle_{AB} = \sqrt{p} |+\rangle |0\rangle + \sqrt{1-p} |-\rangle |1\rangle $.  Now, the coherence of purification is  given by

\begin{equation}
C_{P} (\rho_A) = \min_{U_B} C_{R} (\mathbb{I} \otimes U_B) |\Psi\rangle_{AB}
\end{equation}

\noindent By symmetry, without loss of generality, we choose this unitary to be a real rotation parametrized by angle $\theta$. Hence, the coherence of purification is given by

\begin{equation}
C_{P}(\rho_A) = \min_{\theta} \tilde{H}\left( (\sqrt{p} \cos \theta - \sqrt{1-p} \sin \theta))^2 \right),
\end{equation}

\noindent where $\tilde{H} (x) = - x \log x/2 - (1-x) \log (1-x)/2$. By minimizing over $\theta$ for every $p$, we have to compute the coherence of purification. We note that $\tilde{H}(x)$ is minimized when $x = 0,1$, or in this example, for $\theta = \pm \arccos (\pm \sqrt{1-k})$. By putting this in, we obtain that $C_{P} (\rho_A) = 1$, which is equal to the coherence of purification of the dephased state, which also vindicates our choice of real unitaries, since coherence of purification is monotnous under dephasing maps.


\section{Link with Entanglement of Purification }

One of the motivations behind studying the coherence of purification is supplied by the analogous quantity called the \emph{entanglement of purification} which is a
measure of total correlation for any bipartite state. Given a 
bipartite state $\rho_{AB}$, the entanglement of purification $E_P(AB)$ is defined as \cite{eop}

\begin{equation}
E_P(AB) = min_{A'B'} E_F(\Psi_{AA':BB'}),
\end{equation}
where $\Psi_{AA':BB'}$ is a purification of $\rho_{AB}$ and $E_F(\Psi_{AA':BB'})$ is the entanglement of formation of the pure state across the bipartition 
$AA':BB'$. The entanglement of purification was introduced with the aim that one can define the total correlation of any bipartite state using the entanglement unit in a unified manner. It should be noted that unlike the mutual information which is also a measure of total correlation, the entanglement of purification behaves differently in many respects for tripartite systems \cite{bag}.  Here we show that these two quantities are connected to each other in the following way.

{\bf Proposition 7:}  The coherence of purification $C_P$ of a quantum state $\rho_A$ upper bounds the amount of entanglement of purification $E_P$ that can be created by incoherent operations on the joint system $AB$.

{\bf Proof :} Let us assume the optimal purification for coherence of $\rho_A$ is obtained as $|\Psi \rangle_{AA'}$. That is, $C_P (\rho_A) = C_R ( \Psi_{AA'} )$. Now, there exists an incoherent map $\Lambda_{I}$, such that the relative entropy of entanglement $E_R$ is connected with the relative entropy of coherence $C_R$ in the following manner \citep{Uttam}.

\begin{equation}
E_R(AA' : BB') \left( \Lambda_I [|\Psi\rangle_{AA'} \otimes |0\rangle_{BB'}] \right) = C_R (|\Psi\rangle_{AA'}) = C_P (\rho_A)
\end{equation}
The explicit action of the map $\Lambda_I$ is given by the generalized CNOT unitary $U = \sum_{i=0}^{d_{AA'} -1} \sum_{j=0}^{d_{AA'} -1} |i \rangle_{AA'} \langle i | \otimes |\mod (i+j, d )\rangle_{BB'} \langle j |  + \sum_{i=0}^{d_{AA'} -1} \sum_{j=0}^{d_{BB'} -1} |i \rangle_{AA'} \langle i | \otimes |j \rangle_{BB'} \langle j | $ which is an universal incoherent operation.

Using the fact that for a pure state, $E_F = E_R $ and this purification of the global system $AA':BB'$ may not be optimal from the perspective of entanglement of purification, we have $C_P (\rho_A) = E_{F}(AA':BB') \left( \Lambda_I [|\psi\rangle_{AA'}  \otimes |0\rangle_{BB'}] \right) \geq \min_{A'B'} E_F \left( \Psi_{AA': BB'}\right) = E_{P} (\rho_{AB})$.

Thus,we get the desired result
\begin{equation}
C_P (\rho_A) \geq E_P (\rho_{AB})
\label{lower}
\end{equation}  \qed
This shows that the coherence of purification is the maximum entanglement of purification that can be created via incoherent operation acting on the system and incoherent ancilla. This result complements earlier results that shows how coherence measure can be converted to entanglement measures as well quantum correlation measures. Here, 
even though the coherence of purification is {\em not} a  coherence measure, still it can be converted to total correlation measure under similar setting.


%
%

{\bf Proposition 8 :} The optimal purification of any bipartite quantum state with respect to its coherence of purification is the same as the optimal purification of that state with respect to the entanglement of purification and $C_P(\rho_{AB}) \ge E_P(\rho_{AB})$.

{\bf Proof: } For a bipartite state $\rho_{AB}$ with the corresponding optimal purifications $\Psi_{AA'BB'}$ with respect to the coherence of purification $E_P$. Now, the coherence of purification $C_{P} (\rho_{AB}) = C_{R} (Psi_{AA'BB"}) = C_{R}(\rho_{AA'}) + C_{R} \rho_{BB'} + D(\Psi_{AA'BB'}) $, where $D$ is the symmetric discord, across the bi-partition $AA':BB'$ which equals entanglement entropy for pure states, i.e., $S(AA')$. Therefore,

\begin{equation} C_{P} (\rho_{AB}) \geq C_{R} (\rho_{AA'}) + C_{R} (\rho_{BB'}) + E_{P} (\rho_{AB}),
\label{c1}
\end{equation}
\noindent where the last inequality follows from the fact that the optimal purification from the standpoint of entanglement of purification may be different from that the above purification.

Now, let us assume $|\Phi\rangle_{AA'BB'}$ is the optimal pure state with respect to entanglement purification. then $E_{P} (\rho_{AB}) + C_{R} (\rho_{BB'})+ C_{R} (\rho_{AA'}) = S(\rho_{AA'})+ C_{R} (\rho_{BB'}) + C_{R} (\rho_{AA'})= C_{R} |\phi\rangle_{AA'BB'} \geq C_{P} (\rho_{AB})$.  Therefore,
\begin{equation} C_{P} (\rho_{AB}) \leq C_{R} (\rho_{AA'}) + C_{R} (\rho_{BB'}) + E_{P} (\rho_{AB}),
\label{c2}
\end{equation}

\noindent Combining the inequalities \eqref{c1}, and \eqref{c2}, we note that this is an equality, and hence the optimal purification for $E_{P}$ is also the optimal purification for $C_{P}$ and also $C_P(\rho_{AB}) \ge E_P(\rho_{AB})$. \qed

%
%
%
%
\section{Application to the assisted optimal state discrimination}

In this section, we discuss how coherence of purification has a direct link with the the sucesss probability of the assisted optimal state discrimination (AOSD) protocol \citep{roa, kwek_vedral,  fei_et_al, frey_yoder}-  which seeks to maximise the probability of distinguishing between two different preparations (generally non-orthogonal) of a quantum state in an assisted way, that is, by using an auxillary. This protocol is interesting for the reason that the operational  advantage owing to the utilization of an auxillary persists even when the system and the auxillary qubit do not share any entanglement. It was argued in Ref. \citep{roa} that for the case of zero entanglement, the advantage can be ascribed to the presence of quantum discord, or dissonance, in the joint system-auxiliary state, concentrating specifically on the case where the preparations to be distinguished are equally probable. In this section, we show that in this case, the optimal success probability is solely determined (one-to-one, and monotonically increasing)  by the coherence of purification of the reduced system state. 

\noindent The basic AOSD protocol is the following. A quantum system S is randomly prepared in one of two non-orthogonal states $|\psi_{+}\rangle$, and $|\psi_{-}\rangle$ with a priori probabilities $p_+$, and $p_-$ respectively. The goal is to optimally discriminate between these two preparations. Now, another auxiliary qubit $A$ prepared in the state $|A\rangle$ is coupled to the system by the joint unitary $U_{k}$ such that

\begin{align}
U |\psi_+\rangle |A\rangle = \sqrt{1 - |\alpha_+|^2} |+\rangle |0\rangle + \alpha_+ |0\rangle |1\rangle \nonumber\\
U |\psi_-\rangle |A\rangle = \sqrt{1 - |\alpha_-|^2} |-\rangle |0\rangle + \alpha_- |0\rangle |1\rangle
\end{align}

\noindent It can be easily verified that corresponding overlap function $ \alpha = \langle \psi_{+}|\psi_{-} \rangle = \alpha = \alpha_{+}^{*} \alpha_{-}$ remains unchanged even after the application of the joint unitary. Thus, the prepared joint state of the system and the auxiliary qubit after the joint unitary reads as
\begin{widetext}
\begin{equation}
\rho_{SA} = p_+  U(|\psi_+\rangle \langle |\psi_+| \otimes  |A\rangle \langle A| ) U^{\dagger} + p_-  U  (|\psi_-\rangle \langle |\psi_-| \otimes  |A \rangle\langle A| ) U^{\dagger}
\end{equation}
\end{widetext}

\noindent The corresponding success probability $p_s$ of unambiguous state discrimination is given by \begin{equation}
p_{s} = 1- p_+ |\alpha_+|^2 - p_- |\alpha_-|^2
\end{equation}

To show that the coherence of purification is a resource, we consider the scenario where two states have been prepared 
with equal a priori probabilities, i.e., $p_+ = p_- = 1/2$. In this case the reduced system state $\rho_S$ is given by

\begin{equation}
\rho_S =
\begin{pmatrix}
1- \frac{p_s}{2} & \frac{1}{4} \left( \frac{|\alpha|^2}{|\alpha_+|^2} - |\alpha_+|^2 \right) \\
\frac{1}{4} \left( \frac{|\alpha|^2}{|\alpha_+|^2} - |\alpha_+|^2 \right)  & \frac{p_s}{2}
\end{pmatrix}
\end{equation}

\noindent 

%
%

The concurrence of the joint system-auxiliary state vanishes in the case of \emph{equal a priori probabilities} for the following two conditions \citep{roa}

\begin{enumerate}
\item For optimal success probability, \begin{equation}
|\alpha_+| = \sqrt{|\alpha|}, \hspace{0.3 in} 0 \leq |\alpha| \leq 1
\end{equation}
\item For constant success probability, \begin{equation}
|\alpha_+| = \frac{\sqrt{1 \pm \sqrt{1 - 4 |\alpha|^2}}}{2},  \hspace{0.3 in} 0 \leq |\alpha| \leq 1/2
\end{equation}
\end{enumerate}

In the first case, the systemic density matrix takes the simple form  

\begin{equation}\rho_S =
\begin{pmatrix}
1- \frac{p_s}{2} & 0 \\
0  & \frac{p_s}{2}
\end{pmatrix}
\label{equal_a_priori}
\end{equation}

\begin{figure}
\includegraphics[width = 0.75 \linewidth]{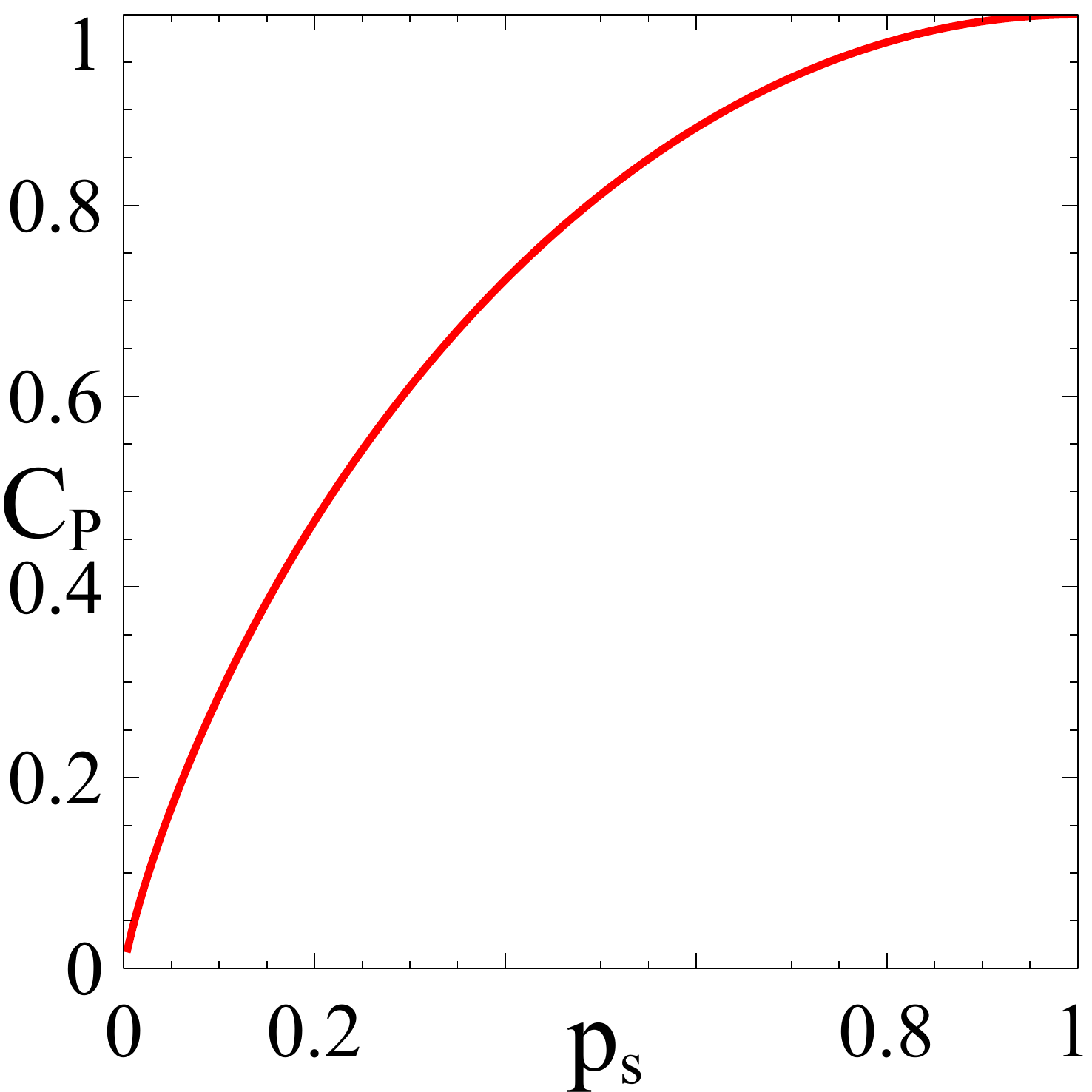}
\caption{Coherence of purification $C_P$vs. success probability $p_s$, in the case of vanishing entanglement and optimal success probability in the AOSD scheme for equal a priori probabilties.}
\label{fig}
\end{figure}

\noindent Thus, the system state $\rho$, given by Eq. \eqref{equal_a_priori} above, is a diagonal state.  For any diagonal state, the coherence of purification is simply the Shannon entropy of its eigenvalues. The eigenvalues here are $\lambda_{1} = 1-\frac{p_s}{2}$, and $\lambda_{2} = \frac{p_{s}}{2}$.  Therefore, the coherence of purification of this state is given by $h(p_s/2)$, where $ h{x} = - x \log_2 x - (1-x) \log_2 (1-x) $ is the binary entropy. Thus, the coherence of purification is always monotonic with the success probability $p_s$ (See Fig. \ref{fig} for a depiction). Therefore, we argue that in the absence of entanglement, the coherence of purification can also be looked at as the resource for assisted optimal state discrimination, as far as obtaining the optimal success probability is concerned.


Interestingly, there is another case where the coherence of purification may again be shown as the operational resource behind the AOSD protocol. Let us imagine that we have an AOSD protocol being executed. Assume that in this setup, the system part of the joint state $\rho_{SA}$ undergoes a completely dephasing channel $\Delta$ \emph{after} the auxiliary system couples with the system qubit. In this case, the coherence of purification of the dephased state $\Delta (\rho_S)$ is formally given once more by the density matrix in Eq. \eqref{equal_a_priori}. Thus, in this case as well, the coherence of purification $C_{P}$ of the completely dephased system state is given by $C_{P} (\Delta(\rho_{S})) = h(\frac{p_s}{2})$,  which once more is monotonous with the success probability. Thus,  in these cases at least, by measuring the coherence of purification one can infer the success probability of the AOSD.

\section{Discussions and Conclusions}

The quest for quantumness in a single system is far from being settled. In this work, we ask: Given a single quantum can there be some quantumness beyond coherence? To answer this question,  we have introduced a quantity
called  the coherence of purification which can be a measure of total quantumness for a single system.
It is defined to be the coherence of the purified state of the density operator with minimum taken over all possible purifications.
The coherence of purification for any mixed state is always larger than the quantum coherence and hence this can capture total quantumness for a single quantum system.
However, the coherence of purification is {\em not} a measure of coherence.
The coherence of purification for a pure state is same the coherence of the state itself, as the state is already a pure state. Therefore, for a
pure state there is nothing beyond quantum coherence.  If we take a
dephased version of the density operator it may still have non-zero coherence of purification. This means that
an incoherent state in a given basis can have finite amount of coherence of purification.
 We have shown that the coherence of purification is the maximal entanglement of purification that can be created using only incoherent operation.
We have also shown that for bipartite states the coherence of purification can be more than the entanglement of purification as well as entanglement of formation.
Finally, we have argued that the coherence of purification is a monotonic function of the success probability in quantum state discrimination problem under certain cases. Our work raises many interesting questions for single as well as multipartite systems. In particular, we note that the regularized entanglement of purification is equal to the  entanglement cost in the limit of asymptotically vanishing classical communication, as proved in Ref. \citep{eop}. In a similar spirit, we conjecture that the regularized version of coherence of purification shall be equal to the coherence cost under some specific conditions. We look forward to a resolution of this conjecture. Furthermore, in future, it will be worth exploring if the coherence of purification acts as useful resource in the absence of quantum coherence, entanglement, quantum discord and other quantum correlations.
	
	\emph{Acknowledgement --} AKP acknowledges Capital Normal University for their kind hospitality. SM acknowledges the support from NSFC under No: 11675113 and 
Beijing Municipal Commission of Education (KZ201810028042). CM acknowledges a doctoral fellowship of Department of Atomic Energy, Govt. of India, and another fellowship by INFOSYS, and thanks Capital Normal University, Beijing, for their kind hospitality.

	\end{document}